# Identifying Brain Image Level Endophenotypes in Epilepsy

Wei Cheng[1,5*], Xuejuan Zhang[1+], Ge Tian[3,5*], Jianfeng Feng[3,4,5], Zhengge Wang[2,5*], Zhiqiang Zhang[2,5], GuangMing Lu[2,5+]

1. Mathematical Department, Zhejiang Normal University, Jinhua, 321004, Zhejiang Provience, P.R. China
2. Department of Radiology, JinLing Hospital of Nanjing, PR China
3. Centre for Computational Systems Biology, Fudan University, Shanghai, 200433, P.R. China
4. Department of Computer Science, Warwick University, Coventry CV4 7AL, UK
5. Fudan University-Jin Ling Hospital Computational Translational Medicine Centre, Fudan University, Shanghai, PR China

**\*: contribute equally**
**+: email:**
xuejuanzhang@gmail.com;        cjr.luguangming@vip.163.com

**Abstract**

A brain wide association study (BWAS) based on the logistic regression was first developed and applied to a large population of epilepsy patients (168) and healthy controls (136). It was found that the most significant links associated with epilepsy are those bilateral links with regions mainly belonging to the default mode network and subcortex, such as amygdala, fusiform gyrus, inferior temporal gyrus, hippocampus, temporal pole, parahippocampal gyrus, insula, middle occipital gyrus, cuneus. These links were found to have much higher odd ratios than other links, and all of them showed reduced functional couplings in patients compared with controls. Interestingly, with the increasing of the seizure onset frequency or duration of illness, the functional connection between these bilateral regions became further reduced. On the other hand, as a functional compensation and brain plasticity, connections of these bilateral regions to other brain regions were abnormally enhanced and became even much stronger with the increase of the seizure onset frequency. Furthermore, patients had higher network efficiencies than healthy controls, and the seizure onset frequency was found to be positively correlated with the network efficiency. A negative correlation between the bilateral connection and the network efficiency was also observed. To further validate our findings, we then employed our BWAS results in discriminating patients from healthy controls and the leave-one-out accuracy was around 78%. Given the fact that a genome-wide association study with a large cohort has failed to identify any significant association between genes and epilepsy, our study could provide us with a set of endophenotypes for further study.

# 1 Introduction

Epilepsy is a brain disorder characterized by an all-encompassing seizure. Currently around 4-10‰ of the population are suffering from the disease, and about 30% of patients cannot be controlled by medications (Forsgren and Hesdorffer, 2009). Different from other brain diseases which have a pure symptom and only a few brain areas are affected (at least at the onset of the disorder), epilepsy is a complicated brain disorder and it has been reported that there are over 40 different subtypes of epilepsy. The source of seizure within brain can be localized (PS=partial seizures) or distributed (GS=generalized seizures), which results in different seizure syndromes and brings difficulties both for medical and clinical treatments (Benbadis, 2001).

Clinical evaluations of epilepsy were previously based on ictal phenomenon, seizure type, syndrome, etiology and impairment. Recently, with the increasing employment of MRI based techniques in assisting clinical diagnosis, classifications of epilepsy are expected to become more objective (Berg *et al.*, 2010, Berg and Scheffer, 2011). Along this line, task-free resting-state fMRI has received considerable interests due to its noninvasive and easiness to execute on patients. By calculating the temporal correlation between fMRI BOLD signals from distinct brain regions, resting-state networks (RSNs) were then derived to represent functional connectivity networks (Bullmore and Bassett, 2011). From a network perspective, human brain is hierarchically organized as a small-world topology in order to simultaneously process information both locally and globally (Bassett and Bullmore, 2006, Achard and Bullmore, 2007). Disruptions of such an optimized functional network may underlie the pathophysiological mechanisms of many psychiatric and neurological diseases, such as Alzheimer (Lo *et al.*, 2010), schizophrenia (Tan *et al.*, 2006, Liu *et al.*, 2008, Zalesky *et al.*, 2011), attention-deficit/hyperactivity disorder (ADHD) (Tomasi and Volkow, 2011), epilepsy (Liao *et al.*, 2010) etc.

In previous studies on epilepsy, altered networks in patients are often observed. For example, default mode network (DMN) was repeatedly reported to be widely affected in epilepsy patients (Luo *et al.*, 2011a, Laufs *et al.*, 2007). In addition, language network, attention network and perceptual network were all found to be altered in such a kind of epilepsy. By exploring dysfunctional connectivity links in intra- and inter RSNs in both temporal lobe epilepsy (TLE) and mixed partial epilepsy (MPE), Luo et al. reported that several RSNs including DMN, dorsal attention network and perceptual network were consistently altered in both TLE and MPE (Luo *et al.*, 2012). Actually, such a consistence was not only found in different kinds of PS, but was also found in GS, although PS patients were generally thought to only have local abnormalities while GS patients do not. Indeed, DMN was also found to be altered in GS patients with idiopathic generalized epilepsy (IGE) (Zhang *et al.*, 2011). Furthermore, both IGE patients and TLE patients showed significant increases in nodal topological characteristics in mesial frontal cortex, putamen, thalamus and amygdala relative to controls (Zhang *et al.*, 2011). All these indicated that although different types of epilepsy have different initialization sources and symptoms, they could exhibit common abnormalities in several brain regions. However, a recent study with 3445 patients and 6935 controls of European ancestry with partial epilepsy has failed to identify common variants (Kasperavičiūtė *et al.* , 2010). Our approach here, to find the common and significant alterations of the functional network in different types of epilepsy, could be very challenge but, if it is successful, it could serve as endophenotypes in future association studies.

It should be pointed out that most results in the literature were mainly based upon a small population of patients (in the order of tens) and the obtained results cannot be fully validated, and hence their clinical significance is still limited. Here we worked on a set of data of large sample size including 168 epilepsy patients but with unknown types and 136 healthy controls. Different from previous studies which usually only applied graph theory analysis, here we first implemented an association study by applying a logistic regression to functional links to find the most significant links related to epilepsy. For these significantly altered links, we further applied graph theory analysis to see whether information processing on these associated regions was also significantly changed in epilepsy patients. Subsequently, we studied how the frequency of seizure onset or duration of illness affects the functional strength of the connectivity network. It was found that both functional connections and network efficiencies were correlated with clinical variables. Finally, a discrimination task was performed to further validate our results and an accuracy of 78.3% with leave-one-out cross validation was achieved.

## 2 Material and Methods

### 2.1 Participants

Participants include 168 patients with epilepsy of unknown types and 136 age- and gender-matched healthy volunteers were recruited as controls from Jinling Hospital Nanjing university school of Medicine.

### 2.2 Data Acquisition

All data were collected on a 3 Tesla Siemens Trio Tim scanner with an eight channel phased array head coil. Resting state fMRI data were acquired axially by using an echo-planar imaging (EPI) sequence. The following parameters are used: TR/TE = 2000ms/30ms, FA = 90°, matrix = 64 ×64, FOV = 24 ×24cm$^2$, and slice thickness =4 mm, slice gap = 0.4mm. A total of 30 slices were used to cover the whole brain. Each section contained 250 volumes. Subjects were instructed to relax, hold still, keep their eyes closed without falling asleep, and think of nothing in particular. Routine anatomical MRI data were acquired to detect structure abnormality.

### 2.3 Data Preprocessing

Data preprocessing was performed using the DPARSF software (Chao-Gan and Yu-Feng, 2010). After slice-timing adjustment and realignment for head-motion correction, the standard Montreal Neurological Institute (MNI) template provided by SPM2 was used for spatial normalization with a resampling voxel size of $3 \times 3 \times 3mm^3$. There were no participants with movement greater than 1*mm* of translation or 1° of rotation. There were also no significant differences between the total range of movement across any axis of translation or rotation between groups.

After smoothing (*FWHM=8mm*), the imaging data were temporally filtered (band pass, 0.01~0.08Hz) to remove the effects of very low-frequency drift and high-frequency noises (e.g., respiratory and cardiac rhythms) (Vincent *et al.*, 2007, Biswal *et al.*, 1995).

## 2.4 Construction of whole-brain functional networks

An automated anatomical labeling (AAL) atlas (Tzourio-Mazoyer *et al.*, 2002) was used to parcellate the brain into 90 regions of interest (ROIs) (45 in each hemisphere). The names of the ROIs and their corresponding abbreviations are listed in Table 1. To rule out the possibility that significant correlation between two regions is a by-product of their correlations with a third region, partial correlation analysis was carried out. The partial correlation coefficient denoted as $C_{ij}$ between any two ROIs *i* and *j* was defined as the minimum partial correlation coefficient of these two regions conditioning on any third region. This avoids estimating large covariance matrices from limited time points and ensures numerical stability.

## 2.5 Brain-wide Association Study (BWAS)

### 2.5.1 Logistic regression

Similar to genome-wide association study (GWAS), we wanted to find an association between the phenotype of brain disorder and functional brain links. It is naturally to borrow the idea of GWAS approach, where logistic regression was very commonly used. Our approach in associating brain disorder and functional links is therefore termed as Brain-wide Association Study (BWAS). There are two alternatively methods for applying logistic regression to overcome the tricky of small sample size but large number of covariates. One is using a univariable logistic regression where 4005 brain links are independently measured; the other one is using Lasso logistic regression where all brain links are considered in a same model but with a penalty term to solve the problem of over fitting.

In the first method, for each pair of brain link with partial correlation $X_{jk}$, each subject consists of a $(Y_i, X_{jk})$ pair, where $Y_i$ is the phenotype for individual $i$ ($Y_i = 1$ for patients, $Y_i = 0$ for controls), and $X_{jk}$ is the partial correlation between brain regions $j$ and $k$. Let $P_i = E(Y_i | X_{jk})$ be the expected value of phenotype given the partial correlation of brain regions $i$ and $j$. By defining $\text{logit}(P) = \log_e(P/1-P)$, the simple binomial logistic regression has the form

$$\text{logit}(P) = \beta_0 + \beta_1 X_{jk},$$

with $\beta_0$ being the intercept, $\beta_1$ being the expected change in $Y$ perunit change in $X_{jk}$. The regression coefficients $\beta_0$ and $\beta_1$ was estimated by maximum likelihood estimator.

For each pair of link between brain regions $i$ and $j$, we calculate the ensemble-averaged partial correlation $\overline{\rho}_{ij}^P$ and $\overline{\rho}_{ij}^H$ for patients group and controls group, respectively. The probability of being a patient for a given partial correlation, the odd for $X_{ij}$ taking value of $\overline{\rho}_{ij}^P$ is $Odd_P = \dfrac{p(Y=1|X_{ij} \leq \overline{\rho}_{ij}^P)}{1 - p(Y=1|X_{ij} \leq \overline{\rho}_{ij}^P)}$, and for $X_{ij}$ taking value of $\overline{\rho}_{ij}^H$ is $Odd_H = \dfrac{p(Y=1|X_{ij} \leq \overline{\rho}_{ij}^H)}{1 - p(Y=1|X_{ij} \leq \overline{\rho}_{ij}^H)}$. The odd ratio for a given brain link is then defined as the ratio between these two odds, i.e., $OR = \dfrac{Odd_P}{Odd_H}$, which is equivalent to $e^{\beta_1(\overline{\rho}_{ij}^P - \overline{\rho}_{ij}^H)}$.

### 2.5.2 Scores of brain links

We endow a score, called generalized risk difference $S_{ij}$ for each link between regions $i$ and $j$ as the difference of the mean partial correlation between healthy controls and patients, i.e.,

$$S_{ij} = \frac{1}{N_P}\sum_{k=1}^{N_P} C_{ij}^P(k) - \frac{1}{N_H}\sum_{k=1}^{N_H} C_{ij}^H(k).$$

$S_{ij}>0$ means that the functional link between regions $i$ and $j$ is decreased in patients compared with controls; while $S_{ij}<0$ means that the coupling between brain regions $i$ and $j$ are abnormally increased in patient. Note that if we use 0 or 1 in the definition above for $C_{ij}^P$ and $C_{ij}^H$, i.e. threshold the network, then $S_{ij}$ is exactly the risk difference.

For the links with $p<0.05$ in regression analysis, we grouped them into two sets: one is composed of links with positive scores, i.e., $S_{Pos} = \{i \leftrightarrow j \mid S_{ij} > 0\}$; and the other is composed of links with negative scores, i.e., $S_{Neg} = \{i \leftrightarrow j \mid S_{ij} < 0\}$. For set $S_{Pos}$, average partial correlation, denoted as $\overline{\rho}_{Pos}$ over these links was calculated for each patient. Similarly, for set $S_{Neg}$, we got $\overline{\rho}_{Neg}$ for each patient.

### 2.6 Network analysis

Each functional connectivity network has a graphic representation, with each node corresponding to an ROI and each link corresponding to a functional connectivity. Let S denote the set of all nodes in the network, $L$ and $R$ be the set of nodes corresponding to ROIs in the left or right hemisphere, respectively; $N_L$ and $N_R$ be the total number of nodes in the left and right hemisphere, respectively. The link connecting the nodes $i$ and $j$ is associated with a weight $w_{ij}$ which in this study is defined as $|C_{ij}|$, i.e. the absolute value of the partial correlation coefficient between *i* and *j* as defined above, to account for the connectivity strength.

The nodal strength of a node *i* is defined as:

$$S_{nodal}(i) = \frac{1}{N-2} \sum_{j \neq i, i_s} |C_{ij}|$$

where $C_{ij}$ is the correlation coefficient between node *i* and node *j*. $i_s$ denote symmetric region of node *i*.

The shortest path length between a pair of nodes *i* and *j* is defined as follows:

$$d_{ij} = \min_{p_{i \leftrightarrow j}} \left\{ L_{ij}^{p_{i \leftrightarrow j}} \right\} = \min_{p_{i \leftrightarrow j}} \left\{ \sum_{(u,v) \in p_{i \leftrightarrow j}} \frac{1}{|C_{uv}|} \right\}$$

where $p_{i \leftrightarrow j}$ is any path that connecting the nodes *i* and *j*, $L_{ij}^{p_{i \leftrightarrow j}}$ is the weighted path length of $p_{i \leftrightarrow j}$, higher correlations are interpreted as shorter distances. Note that $d_{ij} = \infty$ for all disconnected pairs *i, j*.

For a brain network, since it is composed of two symmetric subnetworks belonging to left and right hemispheres, respectively. The left nodal efficiency is defined as:

$$E_i^L = \frac{1}{N_L(N_L - 1)} \sum_{j \in L, j \neq i} d_{ij}^{-1}$$

where $N_L$ is the number of regions in left hemisphere, if *i* belongs to the left hemisphere. Similarly, the right nodal efficiency $E_i^R$ can be defined.

The left and right hemispheric global efficiency of the network are then defined as:

$$E_{glob}^L = \frac{1}{N_L} \sum_{i \in L} E_i^L,$$

$$E_{glob}^{R} = \frac{1}{N_R} \sum_{i \in R} E_i^{R}$$

Finally, the global efficiency of the network is defined as:

$$E_{global} = \frac{1}{2}(E_{glob}^{l} + E_{glob}^{r})$$

## 2.7 Statistical testing

In applying logistic regression in association study, we used Wald test to test whether $\beta_1$ differs significantly from zero. Since the association was tested for a large number of links, a correction for *p*-values was required to account for multiple comparisons. In the present study, a false discovery rate (FDR) correction with significance level *q*<0.05 was used (Storey and Tibshirani, 2003).

Statistical comparisons of characteristics of functional networks such as symmetry, hemispheric nodal efficiency and nodal strength between two groups were performed by using a two-sample two-tailed t test for each value of quantity to determine whether it was significantly different between two groups. Regions where patients and controls show significant difference in hemispheric nodal efficiency were also selected by a FDR correction (*q*<0.05).

Significance test of the correlations of the characteristics of the brain functional networks and the seizure frequency or duration was undertaken by employing a MATLAB code which gives the correlation value and its significance level of difference from zero.

## 3 Results

## 3.1 BWAS in finding the significant links Associated with Epilepsy

We applied the above mentioned BWAS approach to find the most associated links to epilepsy symptom. After the multi-comparison correction (FDR, $q<0.05$), 10 brain links were survived (see Fig. 1). These FDR-survived edges with their ranked *p*-values were presented in Table 2. Interestingly, most of them are links between symmetric brain regions, including bilateral TPOmid, ITG, FFG, HIP which belong to DMN, and AMYG, PHG, MOG and CUN which belong to subcortex, and INS. Besides nine symmetric brain regions, there was a link between PCUN.L and ORBmid.R shown to be significantly altered in patients compared with controls. These FDR survived links with their *p*-values were presented in Table 2. It was seen that the most significantly altered was the one between bilateral AMYG ($p = 5.5 \times 10^{-8}$) which was even survived from the Bonferroni correction; besides, functional connection between bilateral FFG ($p = 6.3 \times 10^{-8}$) was also shown to be survived from this most strict correction.

In Table 2, we further presented the corresponding score values and odds ratios of these ten significant links. It was seen that the scores of the nine links between symmetric brain regions were all positive; while the score of the link between PCUN.L and ORBmid.R was negative. This tells us that for epilepsy patients, functional connections between symmetric brain regions were significantly reduced, which also implies that communication between the left and right hemispheres was hindered in epilepsy patients. Moreover, it was found that the odds ratios of these ten significant links were also ranked as top links among the total 4005 links. It was seen that they are all bigger than 1, i.e. OR>1. Combined with the scores for these significant edges, the fact OR>1 says that for links with positive scores, decreasing the functional connections increases the risk of getting disease; while for links with negative scores, the increase of synchronization between two brain region also increases the risk. This informs us that human brain network was very delicately organized; any unnatural alteration on this network will drive people in a risk of suffering a psychiatric disorder.

It is worthy to note the region AMYG whose bilateral link was found to be the most significantly altered connection. Besides, this region also it was also got the highest score and the highest odds ratio. It was previously reported AMYG was not only altered in PS (Kullmann, 2011, Schreibman Cohen *et al.*, 2009), but also in GS (Zhang *et al.*, 2011). Our findings together with previous reports may provide valuable information that AMYG may be a very high risked region that could be altered in different kinds of epilepsy.

## 3.2 Regional nodal characteristics in epilepsy

Furthermore, we calculated the corresponding hemispheric nodal efficiencies and nodal strength of the above nine pairs of bilateral regions mentioned in Table 2. Comparison of these two quantities between patients and healthy controls was shown in Fig. 2. Interestingly, it was seen that after FDR correction, regions including INS, HIP, PHG, AMYG, CUN, MOG, FFG, TPOmid and ITG which have been mentioned in Table 2, also showed significantly increased hemispheric nodal efficiency in epilepsy patients compared with controls (see Fig. 2 AB). Moreover, nodal strengths of many of these regions such as INS, HIP, PHG, MOG and FFG were also increased in patients, compared with healthy controls ($p<0.05$ uncorrected see Fig. 2 CD). Values of the difference of these two quantities between patients and controls in these brain regions with the corresponding $p$-values were further shown in Table 3.

Furthermore, we calculated hemispheric global efficiency and found that epilepsy patients showed significantly higher hemispheric global efficiency than normal controls in both the left and right hemispheres, with $p$-values $4.6 \times 10^{-5}$ and $7.5 \times 10^{-6}$, respectively.

### 3.3 Correlation with clinical variables

Importantly, we found that characteristics of functional networks such as symmetry, hemispheric nodal efficiency and nodal strength were correlated with clinical variables. As shown in Fig. 3A and B, there is significant correlations between the average partial correlation of 45 pairs of symmetric regions and clinical variables such as seizure frequency ($p=0.0627$) and illness duration ($p=0.0058$). Moreover, for these links belonging to set $S_{Pos}$, i.e. links with positive scores, the corresponding average partial correlation $\bar{\rho}_{Pos}$ was shown to have a significantly positive correlation with the seizure frequency ($p=0.0001$) as well as the illness duration ($p=0.012$). On the other hand, for these links belonging to set $S_{Neg}$, $\bar{\rho}_{Neg}$ was shown to have a significantly negative correlation with the seizure frequency ($p=0.0007$) and the illness duration ($p=0.02$) (see Fig. 4). Further statistical analysis on individual symmetric regions showed that partial correlations of bilateral HIP, PHG, CUN and TPOmid, which all had lower symmetry in patients than in controls, were also shown to be positively correlated with the illness duration (see Fig. 5).

We further investigated the correlation between the network characteristics and the seizure frequency. It was shown in Fig. 3C and D that both global efficiency and mean nodal strength are positively correlated with the seizure frequency ($p=0.0081$ for global efficiency, $p=0.0064$ for mean nodal strength). Furthermore, individual nodal efficiencies of AMYG, FFG, INS, ITG and HIP which were found to be significantly altered in patients were also shown to be positively correlated with seizure frequency (see Fig. 6). These facts revealed that increasing the frequency of seizure onset will result in an increase of global network efficiency as well as individual nodal efficiencies of some important brain regions which may be greatly associated with epilepsy.

## 3.4 Correlation between network efficiency and brain asymmetry

From Fig.1, we have learned that the functional brain networks of epilepsy patients became less synchronized between left and right hemispheres than that of controls. Decoupling the links between symmetric regions led the two hemispheres become more asymmetric. This desynchronization or asymmetry became even more sever with the increase of seizure frequency. On the other hand, the network efficiency of patients was higher than that of controls. We further wanted to know whether the increase of network efficiency was caused by the decrease of functional coupling between the left and right hemispheres. To explore this, we further calculated the correlation between brain network efficiency and the coupling strength of symmetric brain regions. It was seen in Fig.8 that either for patients or controls, the network efficiency was negatively correlated with the coupling strength of symmetric regions. More interestingly, we further found that both hemispheric nodal efficiency and nodal strength are negatively correlated with nodal symmetry as illustrated in Fig.8.

The negative correlation between decreased brain symmetry and increased network efficiencies may imply that although the two hemispheres become less coordinated in patients, local regions tend to connect more closely to each other, as a functional compensation for the reduced coupling between the two hemispheres. To verify this, we had a look of the eight significant pairs of symmetric regions presented in Table 2. In Fig. 7B, we depicted the average coupling strengths of these 16 significant regions to others vs. the nodal symmetry. We found that this is indeed the case.

Therefore we could understand the increase of network efficiency with seizure frequency as a consequence of the frequent appearance of abnormal discharge in epilepsy patients during the periods of seizure onset, which destroys the functional connection between two symmetric brain regions. On the other hand, the obstruction of communication between the two hemispheres established links of other regions to these symmetric regions as a functional compensation. This is also the reason why we saw connections between asymmetric regions enhanced, and hemispheric nodal efficiencies of many regions and global efficiency were significantly increased in patients compared with healthy controls, and become even higher with the increase of seizure frequency.

### 3.5 Application of BWAS in Discrimination

Finally, we applied the above presented BWAS to see to what extent it can classify epilepsy patients from healthy controls. In extracting features selected by BWAS for discrimination, we directly kept those links with *p*-values of Wald test less than 0.01 in the logistic regression. It should be mentioned that these selected links for a feature may change as the subjects in training set changes. The classifier we adopted here is the support vector machine (SVM) implemented with libsvm version 3.1 (Chang and Lin, 2011). To estimate the classification accuracy, we performed the test of SVM using different training and testing data. Only using the feature selected by BWAS, we found that with 60% chosen training data, we have 72% prediction accuracy; with 80% chosen training data, the prediction accuracy is 75%; and with leave-one-out cross-validation (LOOCV), the accuracy reaches 78.3% (see Fig. 9).

For generality, we further calculated the prediction accuracy for different *p*-valve thresholds in selecting featured edges from BWAS, it was seen that for LOOCV, we have more than 70% accuracy of prediction for a wide-range of p-value threshold. For *p*-value thresholds between 0.008~0.02, the accuracy can reach around 78%. Thus the feature extracted from the BWAS approach is very promising in classifying patients and controls. Indeed, it is easy to reach more than 80% prediction accuracy by including other features such as ALFF, ReHo, etc. we would not go further detailed investigation on this.

Furthermore, the method of BWAS in discrimination has its generality in classifying other brain disorders. We applied the same BWAS approach in ADHD dataset with 102 patients and 134 healthy controls. We found that the prediction accuracy of LOOCV could reach 78%.

# 4 Discussions

In the present study, we applied a new method which is similar to GWAS approach, called BWAS here, to find association between brain functional links and epilepsy phenotype. To further understand how brain functional network is altered in patients compared with healthy controls, we further applied the graph theory to analyze the network efficiency in epilepsy patients. Our major findings are: i). In general, epilepsy patients had reduced functional coupling between left and right hemispheres, which says that brain symmetry was reduced in patients, compared with controls. ii). The network efficiencies including global efficiency and nodal efficiencies of bilateral brain regions were significantly increased in patients compared with controls. iii). The altered network characteristics were significantly correlated with clinical variables, such as seizure onset frequency and illness duration. In particular, the global and nodal efficiencies were positively correlated with seizure frequency; while the reduced brain symmetry was negatively correlated with both seizure frequency and illness duration. iv). Moreover, global network efficiency and the coupling strength between symmetric regions was negatively correlated.

## 4.1 Association of disordered brain networks and psychiatric diseases

Using functional connectivity analysis of low-frequency BOLD signal in resting state, many brain disorders have been revealed to have altered functional connections of brain network. Here we applied a whole brain association study to systematically assess whether a functional link between two brain regions contributes to the risk of epilepsy. After a strict FDR correction ($q<0.05$), we found that the most significant changes in functional circuit of epilepsy patients were mainly occurred in symmetric brain regions. Interestingly, these links of these symmetric regions all have reduced functional coupling in patients compared with controls.

The regression analysis on association study could also be reliably applied to other datasets of brain disorders. Here we took ADHD for verification. For ADHD, using the permutation method, it has been shown in (Ji *et al*., 2012) that the most significantly altered functional links associated to ADHD was the increased coupling of the saliency network involving the anterior cingulate gyrus and anterior insula. Our regression approach also reproduced this salience network.

The association study of the functional connectivities and different mental diseases proposed that different psychiatric diseases may be characterized by a specific alteration in brain network organization: For ADHD, saliency network may represent a remarkable disruption associated to this particular disorder in children. For epilepsy, the most significant alteration was characterized by a significant reduction of functional couplings between symmetric regions which are mainly belonging to DMN and subcortex.

## 4.2 Alteration of network properties

Epilepsy is one of the most complicated brain disorders in which almost all brain regions could be involved. Previous studies mainly focused on pure types of epilepsy but only with a small sample size. Several studies on resting fMRI of temporal lobe epilepsy, a particular type of PS, have uncovered decreased functional connections in DMN, impaired perceptual network, language network were also revealed in temporal lobe epilepsy (Luo *et al.*, 2011a). On the other hand, recent studies on idiopathic generalized epilepsy, found that DMN was also disrupted in this type of GS patients (Zhang *et al.*, 2011).

Although the patients we analyzed were of unknown types of epilepsy, the results we revealed are based on a large sample size and are still in consistent with previous findings. Firstly, we also found decreased coupling within DMN, where HIP, ITG and TPOmid were all to have showed reduced couplings within their bilateral parts in patients. On the other hand, PCUN.L which is also in DNM showed an enhanced coupling with ORBmid.R, this may serve as a concomitant compensation. The region PCUN was further shown to have increased hemispheric nodal efficiency in left and right parts.

Secondly, our association study showed that the most significantly changed regions was AMYG. This region was shown to have a reduced coupling between left and right parts, but with an increased nodal efficiency. Actually, alterations in AMYG were consistently reported in PS such as temporal lobe epilepsy (Bonelli *et al.*, 2009, Mitsueda-Ono *et al.*, 2011), as well as GS such as IGE (Zhang *et al.*, 2011). Using K-means method, our previous study on a less small sample size with 100 patients and 80 controls also showed that AMYG was the most significant region (Zhang *et al.*, 2012).

Thirdly, Besides AMYG, there are many regions in subcortex that were found to have decreased couplings between symmetric regions, such as FFG, PHG, MOG.

Among these regions, the connection between bilateral FFG was also survived form the Bonferroni correction ($p$=1.1e-7). FFG is part of the temporal lobe. It is involved in face perception. It was reported that attention and perception deficit is a common symptom in temporal lobe epilepsy (Zhang *et al.* , 2009), and recent study on partial epilepsy showed that both temporal lobe epilepsy and mixed partial epilepsy have decreased functional network connection in dorsal attention network and perceptual network (Luo *et al.* , 2012). Recent study on white matter showed that temporal lobe epilepsy patients have significant white matter reduction in the bilateral HIP, PHG and FFG. Reduced functional coupling between left and right HIP was also reported in partial seizure patients (Firings et al., 2009).

## 4.3 Correlation with clinical variables

More importantly, we found that the functional links selected by BWAS were correlated with clinical variables. Our results revealed that the connection strength between symmetric brain regions is negatively correlated with clinical variables. This says that increasing the seizure frequency most probably hinders the communications between left and right hemispheres. As a functional compensation, connection between other brain regions may be abnormally enhanced. These abnormally enhanced links were found to be negatively correlated with clinical variables. This is quite interesting, and has not been reported by previous papers. Furthermore, our analysis revealed that for those regions with reduced coupling in patients, increasing seizure frequency or illness duration will further disrupt the coupling between them; while for those brain regions with enhanced coupling in patients, increasing seizure frequency or illness duration of ill will further abnormally enhance the synchronization between them. This might explain why the symptoms of epilepsy patients become more and more sever with the increase times of seizure onset.

Furthermore, by applying graph theoretical analysis, we found that patients have significantly higher global efficiency than controls, and this network efficiency is positively correlated with seizure frequency. Both epilepsy patients and controls have a small-network structural connectivity, but patients have an aberrant small-worldness. But whether the functional network in epilepsy patents transits toward a more regular or toward a more random structure is still in a debate: Some results indicated a trend towards a more regular shift (Horstmann *et al.*, 2010), while others showed a random shift (Liao *et al.*, 2010, Schindler *et al.*, 2008, Chavez *et al.*, 2010, Kramer *et al.*, 2010). The increase of both global and nodal efficiencies in our datasets consistently supported the tendency towards a random network configuration. Combined our result observed from mixed types of epilepsy with previous studies on both GS (Zhang *et al.*, 2011) and PS (Liao *et al.*, 2010), it seems that the increase of seizure onset had a same consequence of increasing the network efficiency on GS and PS patients.

The increase of network efficiency with the increase of seizure can be understood from the relationship between global network efficiency and the coupling strength between left and right hemispheres. Actually, decreasing the coupling strength between symmetric regions increased the possibility of their connections to others. And this mechanism is true both for healthy controls and patients. Therefore, both of them showed a negative correlation between these two quantities.

## 4.4 Towards an automatic classification of epilepsy

A vast majority of fMRI studies on psychiatric diseases only reported altered brain network in psychiatric patients by undertaking group comparison, it is still unknown whether these altered brain network has clinical significance.

In the current paper, we further developed a classifier using the BWAS method mentioned above to see whether it can discriminate epilepsy patients from healthy controls successfully. Here using only a single classifier based on BWAS method, we can successfully classify patients from controls with 78% accuracy using LOOCV. And this method was found to be general to other mental disease such as ADHD, which can reach 78% prediction accuracy using LOOCV. We therefore anticipated the BWAS-based classifier may be served as an automatic classification for mental disease.

Although we have achieved a high accuracy of discrimination, we have much space for further improvements. Here we have only employed the fMRI data; structural MRI data which should be informative has not been included yet. On the other hand, we have only used information from BOLD signals here and have not taken into account T1-weighted signals and other modalities such as diffusion tensor images etc. (Wee *et al.*, 2011, Dai *et al.*, 2011). Even with BOLD signals, we can include effective networks as further features (Luo *et al.*, 2011b, Zou and Feng, 2009, Ge *et al.*, 2011). In summary, to achieve a reliable and fast diagnosis of various brain disorders, we have to take up a multi-modal approach.

# Acknowledgement

JF is a Royal society Wolfson research merit award holder.

# References

Forsgren L, Hesdorffer D. Epidemiology and prognosis of epilepsy. The Treatment of Epilepsy. 2009: 21-31.

Benbadis SR. Epileptic seizures and syndromes. Neurologic clinics. 2001; 19(2).

Berg AT, Berkovic SF, Brodie MJ, Buchhalter J, Cross JH, Van Emde Boas W, et al. Revised terminology and concepts for organization of seizures and epilepsies: report of the ILAE Commission on Classification and Terminology, 2005–2009. Epilepsia. 2010; 51(4): 676-85.

Berg AT, Scheffer IE. New concepts in classification of the epilepsies: Entering the 21st century. Epilepsia. 2011.

Bullmore ET, Bassett DS. Brain graphs: graphical models of the human brain connectome. Annual review of clinical psychology. 2011; 7: 113-40.

Bassett DS, Bullmore E. Small-world brain networks. The neuroscientist. 2006; 12(6): 512-23.

Achard S, Bullmore E. Efficiency and cost of economical brain functional networks. PLoS computational biology. 2007; 3(2): e17.

Lo CY, Wang PN, Chou KH, Wang J, He Y, Lin CP. Diffusion tensor tractography reveals abnormal topological organization in structural cortical networks in Alzheimer's disease. The Journal of Neuroscience. 2010; 30(50): 16876-85.

Tan HY, Sust S, Buckholtz J, Mattay V, Meyer-Lindenberg A, Egan M, et al. Dysfunctional prefrontal regional specialization and compensation in schizophrenia. American Journal of Psychiatry. 2006; 163(11): 1969-77.

Liu Y, Liang M, Zhou Y, He Y, Hao Y, Song M, et al. Disrupted small-world networks in schizophrenia. Brain. 2008; 131(4): 945-61.

Zalesky A, Fornito A, Seal ML, Cocchi L, Westin CF, Bullmore ET, et al. Disrupted axonal fiber connectivity in schizophrenia. Biological psychiatry. 2011; 69(1): 80-9.

Tomasi D, Volkow ND. Abnormal Functional Connectivity in Children with Attention-Deficit/Hyperactivity Disorder. Biological psychiatry. 2011.

Liao W, Zhang Z, Pan Z, Mantini D, Ding J, Duan X, et al. Altered functional connectivity and small-world in mesial temporal lobe epilepsy. PloS one. 2010; 5(1): e8525.

Luo C, Li Q, Lai Y, Xia Y, Qin Y, Liao W, et al. Altered functional connectivity in default mode network in absence epilepsy: A resting‐state fMRI study. Human brain mapping. 2011a; 32(3): 438-49.

Laufs H, Hamandi K, Salek‐Haddadi A, Kleinschmidt AK, Duncan JS, Lemieux L. Temporal lobe interictal epileptic discharges affect cerebral activity in "default mode" brain regions. Human brain mapping. 2007; 28(10): 1023-32.

Luo C, Qiu C, Guo Z, Fang J, Li Q, Lei X, et al. Disrupted Functional Brain Connectivity in Partial Epilepsy: A Resting-State fMRI Study. PloS one. 2012; 7(1): e28196.

Zhang Z, Liao W, Chen H, Mantini D, Ding JR, Xu Q, et al. Altered functional–structural coupling of large-scale brain networks in idiopathic generalized epilepsy. Brain. 2011; 134(10): 2912-28.

Kasperavičiūtė D, Catarino CB, Heinzen EL, Depondt C, Cavalleri GL, Caboclo LO, et al. Common genetic variation and susceptibility to partial epilepsies: a genome-wide association study. Brain. 2010; 133(7): 2136-47.

Chao-Gan Y, Yu-Feng Z. DPARSF: a MATLAB toolbox for "pipeline" data analysis of resting-state fMRI. Frontiers in systems neuroscience. 2010; 4.

Vincent J, Patel G, Fox M, Snyder AZ, Baker J, Van Essen D, et al. Intrinsic functional architecture in the anaesthetized monkey brain. Nature. 2007; 447(7140): 83-6.

Biswal B, Zerrin Yetkin F, Haughton VM, Hyde JS. Functional connectivity in the motor cortex of resting human brain using echo‐planar mri. Magnetic resonance in medicine. 1995; 34(4): 537-41.

Tzourio-Mazoyer N, Landeau B, Papathanassiou D, Crivello F, Etard O, Delcroix N, et al. Automated anatomical labeling of activations in SPM using a macroscopic anatomical parcellation of the MNI MRI single-subject brain. Neuroimage. 2002; 15(1): 273-89.

Storey JD, Tibshirani R. Statistical significance for genomewide studies. Proceedings of the National Academy of Sciences of the United States of America. 2003; 100(16): 9440.

Kullmann DM. What's wrong with the amygdala in temporal lobe epilepsy? Brain. 2011; 134(10): 2800-1.

Schreibman Cohen A, Daley M, Siddarth P, Levitt J, Loesch IK, Altshuler L, et al. Amygdala volumes in childhood absence epilepsy. Epilepsy & Behavior. 2009; 16(3): 436-41.

Chang CC, Lin CJ. LIBSVM: a library for support vector machines. ACM Transactions on Intelligent Systems and Technology (TIST). 2011; 2(3): 27.

Bonelli SB, Powell R, Yogarajah M, Thompson PJ, Symms MR, Koepp MJ, et al. Preoperative amygdala fMRI in temporal lobe epilepsy. Epilepsia. 2009; 50(2): 217-27.

Mitsueda-Ono T, Ikeda A, Inouchi M, Takaya S, Matsumoto R, Hanakawa T, et al. Amygdalar enlargement in patients with temporal lobe epilepsy. Journal of Neurology, Neurosurgery & Psychiatry. 2011; 82(6): 652.

Zhang Z, Lu G, Zhong Y, Tan Q, Yang Z, Liao W, et al. Impaired attention network in temporal lobe epilepsy: a resting FMRI study. Neuroscience letters. 2009; 458(3): 97-101.

Horstmann MT, Bialonski S, Noennig N, Mai H, Prusseit J, Hinrichs H, et al. State dependent properties of epileptic brain networks: comparative graph-theoretical analyses of simultaneously recorded EEG and MEG. Clinical Neurophysiology. 2010; 121(2): 172-85.

Schindler KA, Bialonski S, Horstmann MT, Elger CE, Lehnertz K. Evolving functional network properties and synchronizability during human epileptic seizures. Chaos: An Interdisciplinary Journal of Nonlinear Science. 2008; 18(3): 033119--6.

Chavez M, Valencia M, Navarro V, Latora V, Martinerie J. Functional modularity of background activities in normal and epileptic brain networks. Physical review letters. 2010; 104(11): 118701.

Kramer MA, Eden UT, Kolaczyk ED, Zepeda R, Eskandar EN, Cash SS. Coalescence and fragmentation of cortical networks during focal seizures. The Journal of Neuroscience. 2010; 30(30): 10076-85.

Wee CY, Yap PT, Li W, Denny K, Browndyke JN, Potter GG, et al. Enriched white matter connectivity networks for accurate identification of MCI patients. Neuroimage. 2011; 54(3): 1812-22.

Dai D, He H, Vogelstein J, Hou Z. Network-based classification using cortical thickness of AD patients. Machine Learning in Medical Imaging. 2011: 193-200.

**Luo Q, Ge T, Feng J. Granger causality with signal-dependent noise. Neuroimage. 2011b; 57(4): 1422-9.**

**Zou C, Feng J. Granger causality vs. dynamic Bayesian network inference: a comparative study. BMC bioinformatics. 2009; 10(1): 122.**

**Ge T, Feng J, Grabenhorst F, Rolls ET. Componential Granger causality, and its application to identifying the source and mechanisms of the top-down biased activation that controls attention to affective vs sensory processing. Neuroimage. 2011.**

**Table 1.** The names and abbreviations of the regions of interest (ROIs).

| Regions | Abbr. | Regions | Abbr. |
|---|---|---|---|
| Amygdala | AMYG | Orbitofrontal cortex (middle) | ORBmid |
| Angular gyrus | ANG | Orbitofrontal cortex (superior) | ORBsup |
| Anterior cingulate gyrus | ACG | Pallidum | PAL |
| Calcarine cortex | CAL | Paracentral lobule | PCL |
| Caudate | CAU | Parahippocampalgyrus | PHG |
| Cuneus | CUN | Postcentralgyrus | PoCG |
| Fusiform gyrus | FFG | Posterior cingulate gyrus | PCG |
| Heschlgyrus | HES | Precentralgyrus | PreCG |
| Hippocampus | HIP | Precuneus | PCUN |
| Inferior occipital gyrus | IOG | Putamen | PUT |
| Inferior frontal gyrus (opercula) | IFGoperc | Rectus gyrus | REC |
| Inferior frontal gyrus (triangular) | IFGtriang | Rolandic operculum | ROL |
| Inferior parietal lobule | IPL | Superior occipital gyrus | SOG |
| Inferior temporal gyrus | ITG | Superior frontal gyrus (dorsal) | SFGdor |
| Insula | INS | Superior frontal gyrus (medial) | SFGmed |
| Lingual gyrus | LING | Superior parietal gyrus | SPG |
| Middle cingulate gyrus | MCG | Superior temporal gyrus | STG |
| Middle occipital gyrus | MOG | Supplementary motor area | SMA |
| Middle frontal gyrus | MFG | Supramarginalgyrus | SMG |
| Middle temporal gyrus | MTG | Temporal pole (middle) | TPOmid |
| Olfactory | OLF | Temporal pole (superior) | TPOsup |
| Orbitofrontal cortex (inferior) | ORBinf | Thalamus | THA |
| Orbitofrontal cortex (medial) | ORBmed | | |

**Table2. Significant links survived from FDR correction of BWAS approach.**

| Edges | $P$ value | Odds Ratio | Diff. of Partial Corr. |
|---|---|---|---|
| **AMYG.L------AMYG.R** | $5.5\times10^{-8}$ | 1.848 | -0.114 |
| **FFG.L------FFG.R** | $6.3\times10^{-8}$ | 1.792 | -0.104 |
| **TPOmid.L------TPOmid.R** | $2.1\times10^{-5}$ | 1.356 | -0.102 |
| **PHG.L------PHG.R** | $5.4\times10^{-5}$ | 1.276 | -0.093 |
| **MOG.L------MOG.R** | $5.8\times10^{-5}$ | 1.314 | -0.089 |
| **HIP.L------HIP.R** | $5.8\times10^{-5}$ | 1.262 | -0.096 |
| **INS.L------INS.R** | $9.0\times10^{-5}$ | 1.354 | -0.059 |
| **ITG.L------ITG.R** | $1.1\times10^{-4}$ | 1.255 | -0.080 |
| **CUN.L------CUN.R** | $1.2\times10^{-4}$ | 1.400 | -0.073 |
| **ORBmid.R------PCUN.L** | $1.2\times10^{-4}$ | 1.169 | 0.043 |

**Table3.Regional nodal characteristics**

| Regions | Nodal efficiency | | | | Nodal strength | | | |
|---|---|---|---|---|---|---|---|---|
| | Difference | | $P$ value | | Difference | | $P$ value | |
| | Right | Left | Right | Left | Right | Left | Right | Left |
| **INS** | -0.0057 | -0.0115 | 0.015 | $1.3 \times 10^{-6}$ | -0.0012 | 0.0025 | NS | 0.0267 |
| **HIP** | -0.0067 | -0.006 | $3.7 \times 10^{-4}$ | $7.5 \times 10^{-4}$ | -0.0019 | -0.0018 | 0.0244 | 0.0339 |
| **PHG** | -0.0080 | -0.0044 | $2.9 \times 10^{-5}$ | 0.0238 | -0.0024 | 0.0004 | 0.0157 | NS |
| **AMYG** | -0.0053 | -0.0049 | 0.0155 | 0.0169 | -0.0018 | -0.0013 | NS | NS |
| **CUN** | -0.0066 | -0.0057 | $8.1 \times 10^{-4}$ | 0.006 | -0.0011 | -0.0004 | NS | NS |
| **MOG** | -0.0042 | -0.0055 | 0.0205 | 0.0028 | -0.0019 | -0.002 | 0.0406 | 0.0532 |
| **FFG** | -0.0083 | -0.0054 | $1.8 \times 10^{-4}$ | 0.0067 | -0.0035 | -0.0018 | 0.0035 | NS |
| **TPOmid** | -0.0053 | -0.0024 | 0.0139 | NS | -0.0011 | 0.0004 | NS | NS |
| **ITG** | -0.0054 | 0.0004 | 0.0049 | NS | -0.0001 | -0.0008 | NS | NS |

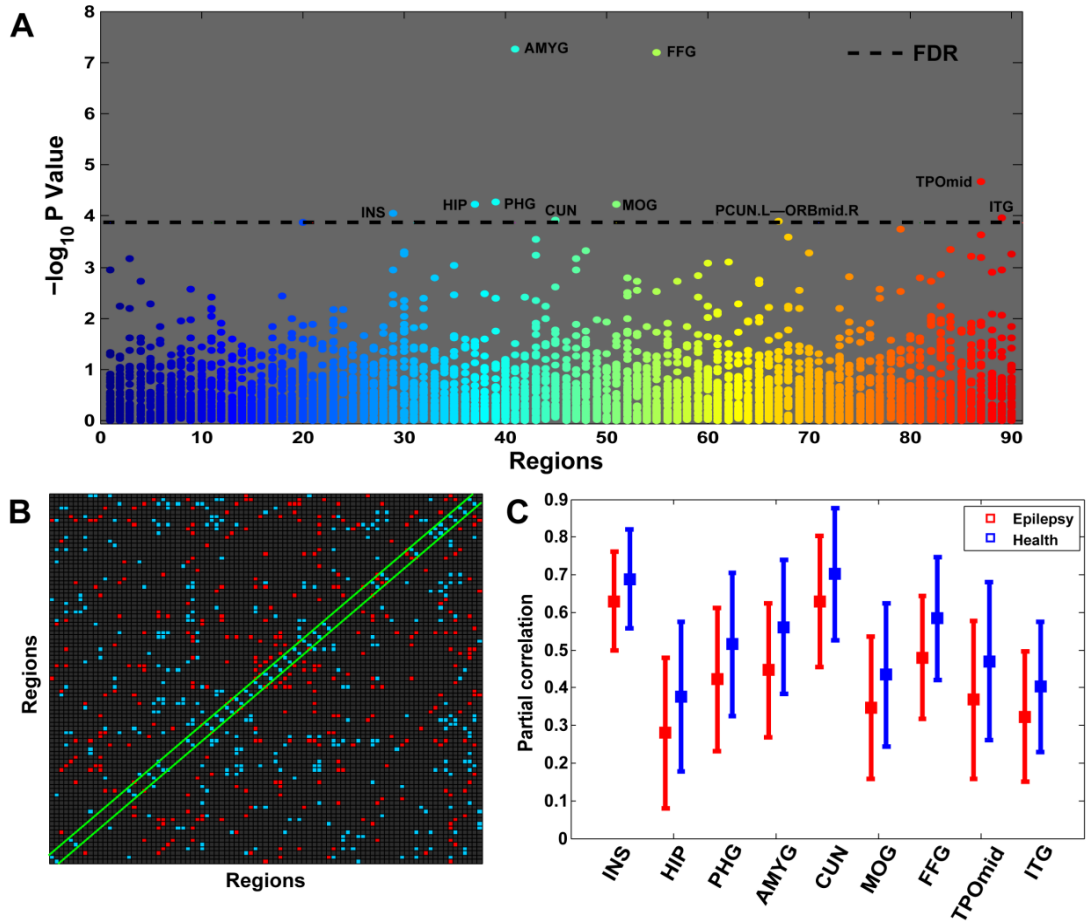

**Figure 1.**(A) *p*-values of edges calculated from the Wald test of the logistic regression, the ones above the dashed line are survived from FDR correction. (B) The significant links which *p*-values are less than 0.05, red dots are links increase in patients' network while blue dots are links decrease in patients' network. (C) Partial correlations of regions that are survived from FDR correction in (A). Red ones are for patients and blue ones are for health controls.

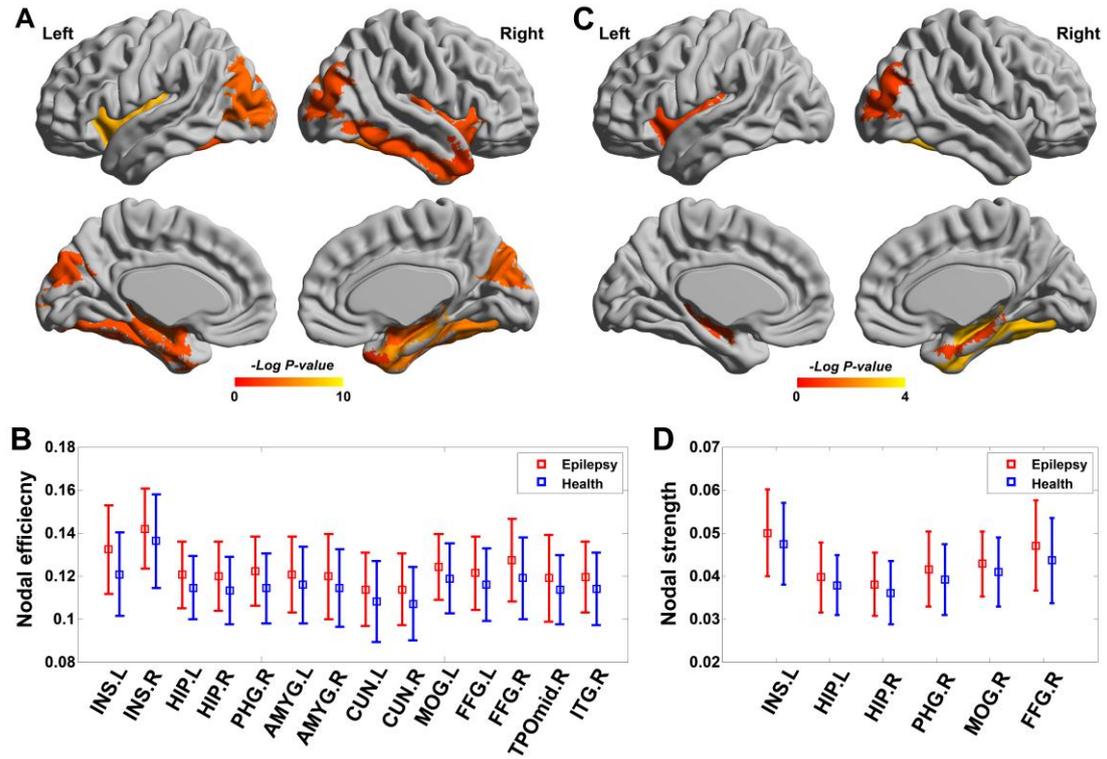

**Figure 2.**(A) (B) Comparison of nodal efficiencies, Regions where patients and control have significant difference in nodal efficiencies (FDR correction, $q<0.05$). The corresponding nodal efficiencies of these significant regions for two groups were shown in (C). Regions that have significantly increased nodal strength in patients (red) than controls (blue) were found to be INS.L, bilateral HIP, PHG.R, MOG.R, FFG.R. the corresponding nodal strength of these regions were shown in (D).

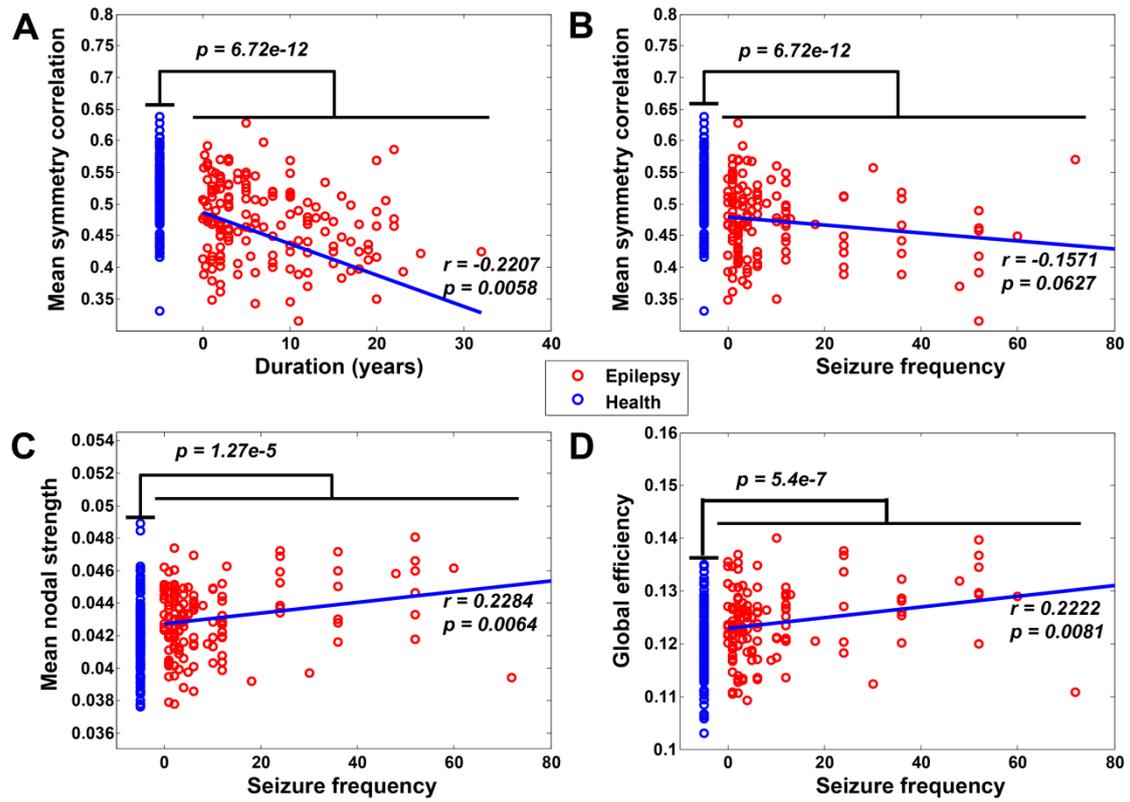

**Figure 3.**(A) (B)Compared to healthy controls, epilepsy patients showed decreased symmetry ($p$=6.72e-12), and the averaged partial correlation over the whole 45 links between symmetric regions is negatively correlated with seizure frequency and the illness duration($p$=0.0627, $p$=0.0058). (C) (D) epilepsy patients showed increased nodal strength ($p$=1.27e-5) and global efficiency ($p$=5.4e-7), and both of them are positively correlated with seizure frequency($p$=0.0064, $p$ =0.0081).

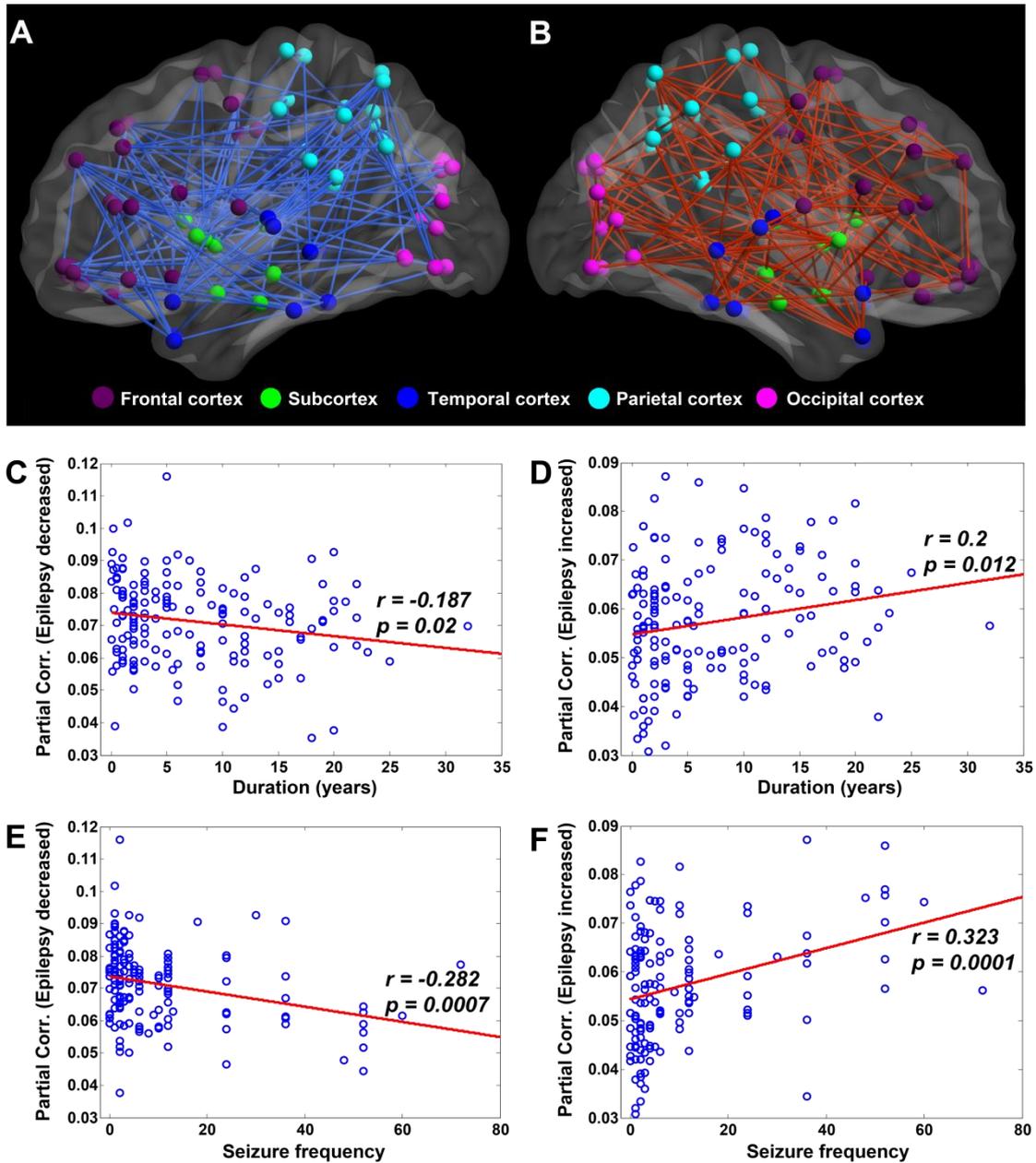

**Figure 4.** (A) Links in set $S_{Neg}$, i.e., the ones that patients have lower scores than controls for $p<0.05$ in the logistic regression. (B) Links in set $S_{Pos}$, i.e., the ones that patients have higher scores than controls for $p<0.05$ in the logistic regression. (C) (E) Scatterplots depicting the correlation between the average partial correlation calculated over links in A and clinical variables. (D) (F) Scatterplots depicting the positive correlation between the averaged partial correlation calculated over links in B and clinical variables.

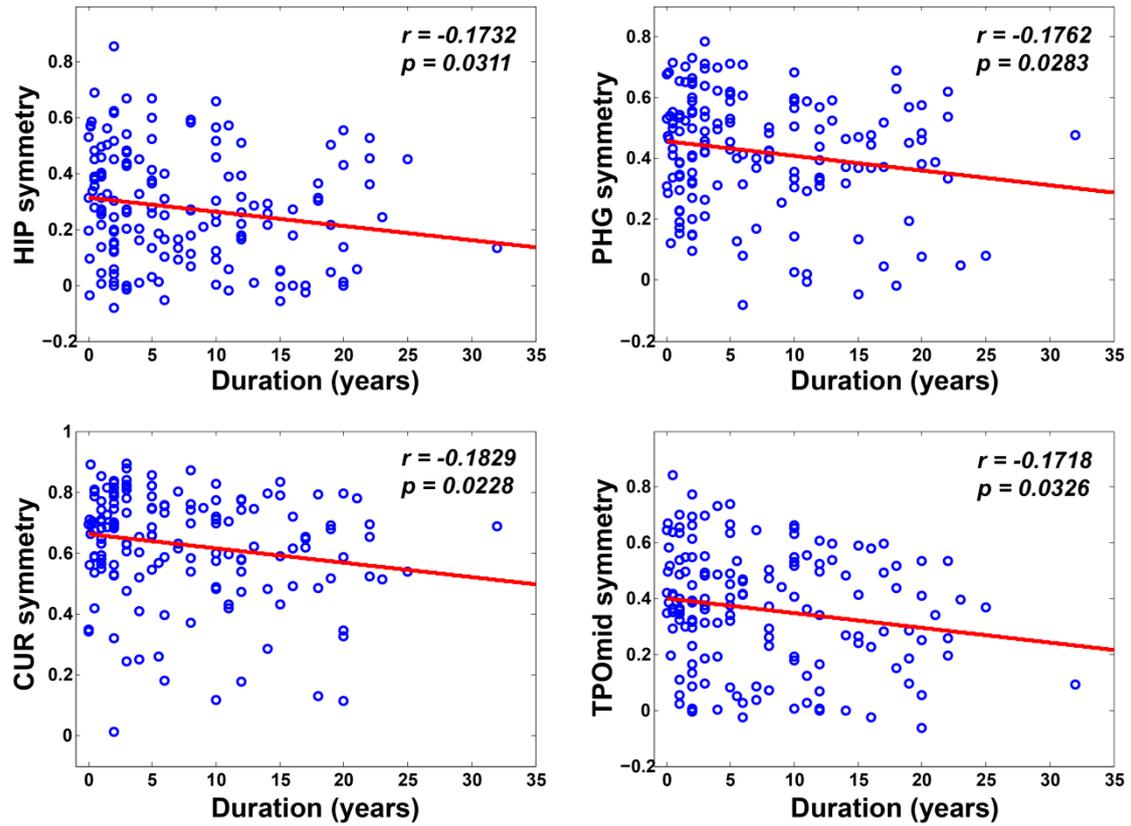

**Figure 5.** Scatterplots of partial correlation of individual pairs of bilateral regions vs. illness duration. It is showed that HIP, PHG, CAU, TPOmid which have lower symmetry in patients than in controls show negative correlation between the bilateral partial correlation and the illness duration.

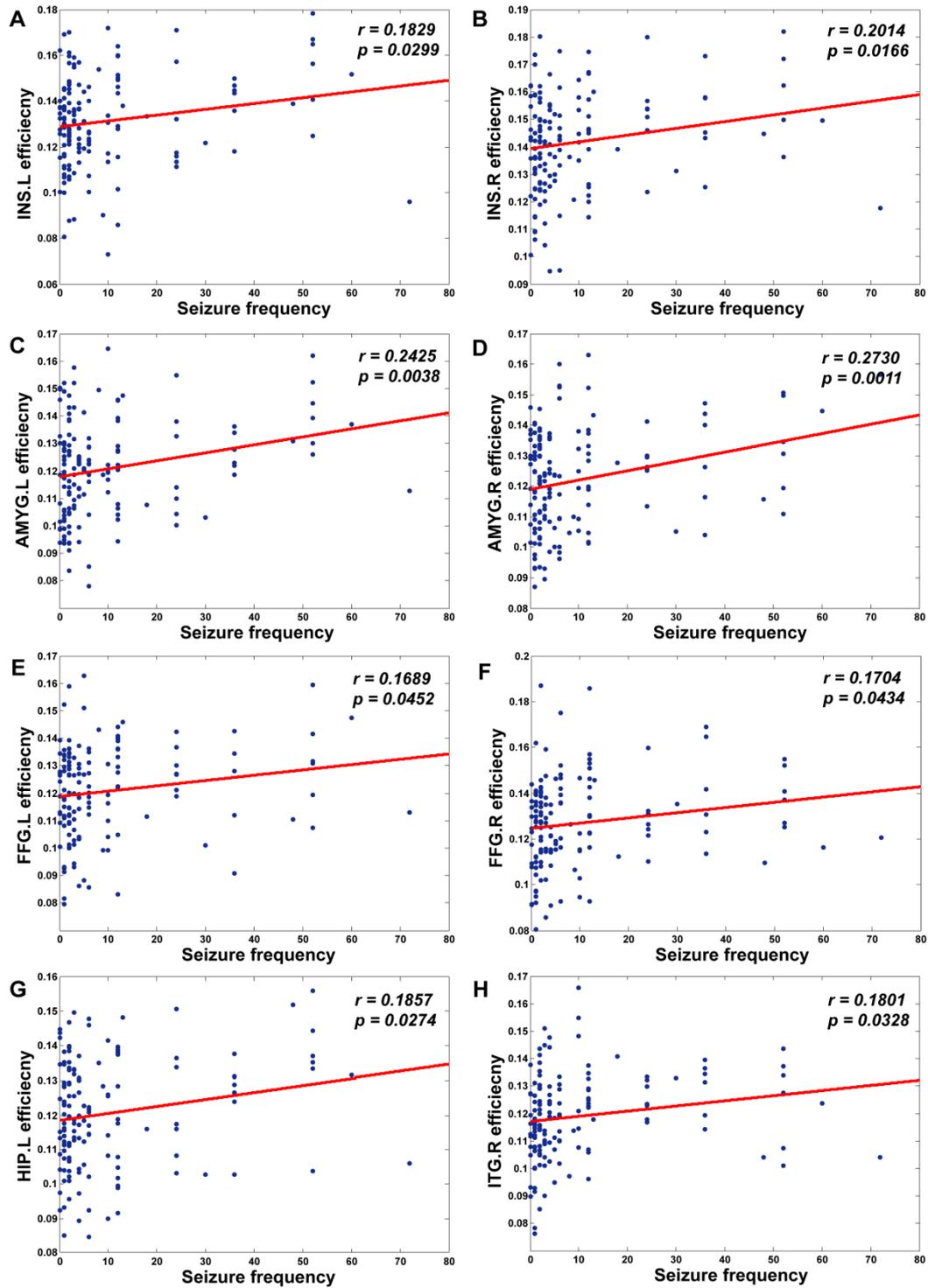

**Figure 6.** Scatterplots hemispheric nodal efficiency and seizure frequency show positive correlation between these two quantities.

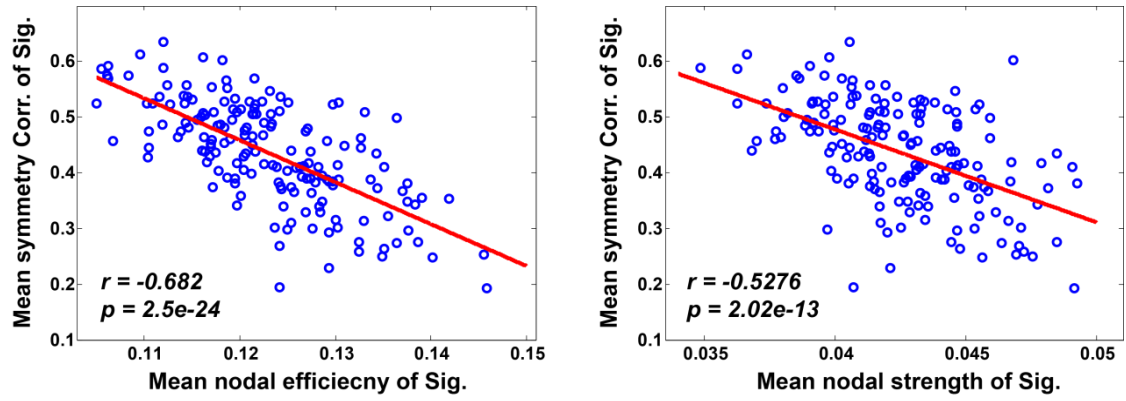

**Figure 7.** Scatterplots of the average connection strength of nine links between symmetric regions vs. hemispheric nodal efficiency (A) and nodal strength (B) of the corresponding regions. It was shown both hemispheric nodal efficiency and nodal strength exhibited negative correlation with nodal symmetry.

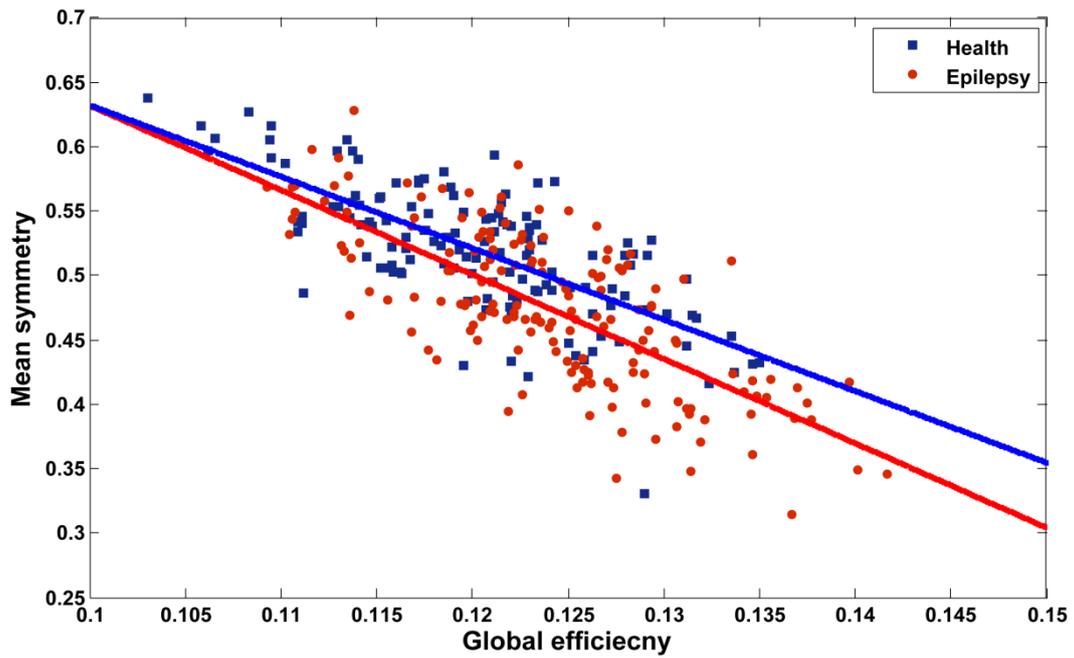

**Figure 8.** Scatterplots of the global network efficiency vs. the average connection strength of the 45 links of symmetric regions. It was shown that both patents and controls exhibited negative correlation between these two quantities.

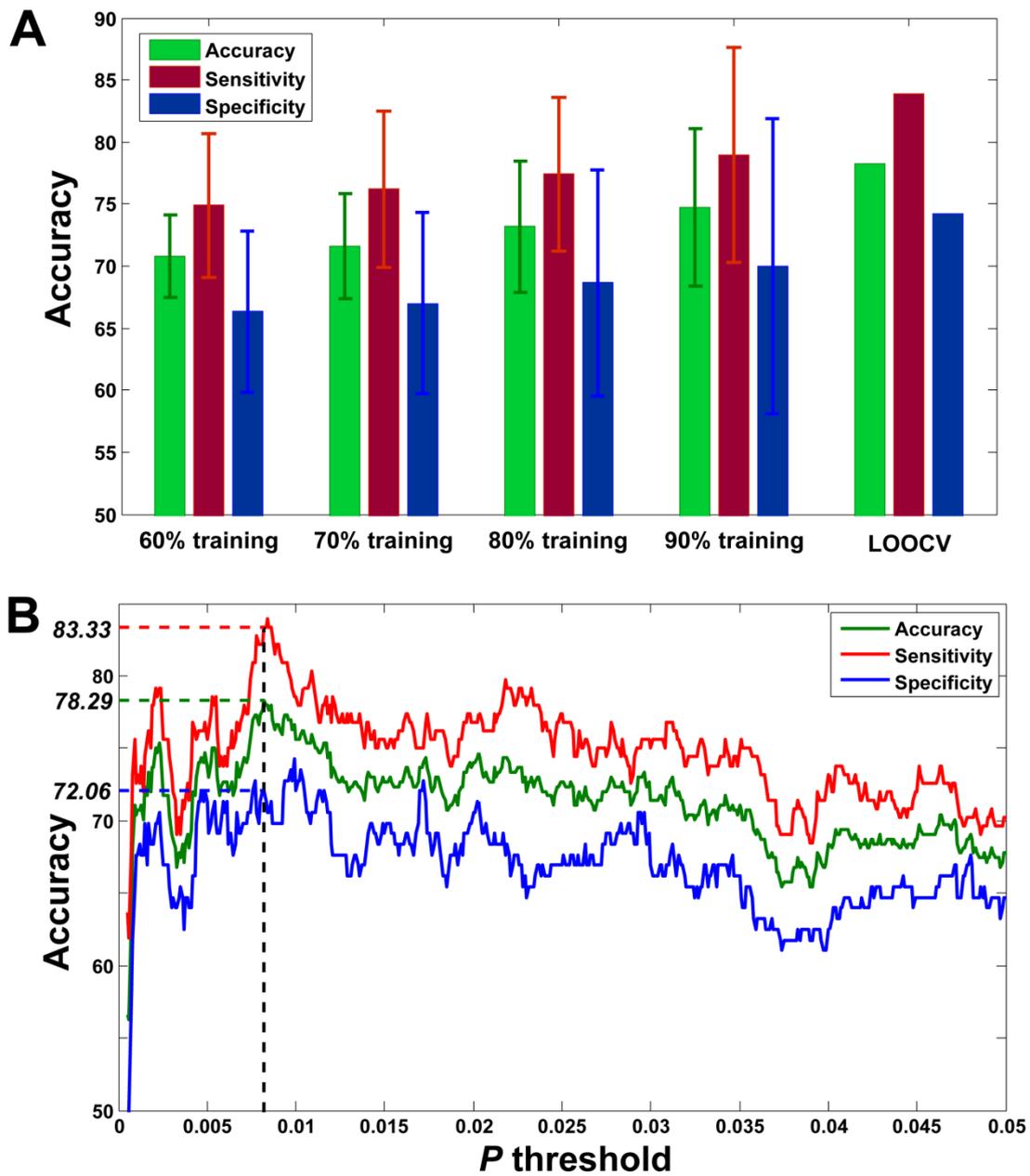

**Figure 9.** LOOCV prediction accuracy of classing epilepsy patients from controls with respect to different *p*-value thresholds in the logistic regression. Corresponding sensitivity and specificity were also plotted. It was seen that for a wide range of *p*-value thresholds in selecting links for SVM, we have more than 70% accuracy of predictions. For *p*-value thresholds between 0.008~0.02, the accuracy can reach around 78%.